\documentclass[a4paper,twocolumn,11pt,accepted=2023-12-14]{quantumarticle}

\pdfoutput=1

\usepackage[utf8]{inputenc}
\usepackage[english]{babel}
\usepackage[T1]{fontenc}
\usepackage{amsmath}
\usepackage{hyperref}
\hypersetup{
    colorlinks=true,       
    linkcolor=cyan,          
    citecolor=magenta,        
    filecolor=magenta,      
    urlcolor=cyan,           
    runcolor=cyan
}

\usepackage{tikz}
\usepackage{lipsum}

\usepackage{mathrsfs}
\usepackage{multirow}
\usepackage{amsmath}
\usepackage{amsthm}
\usepackage{amssymb}
\usepackage{graphicx}
\usepackage{booktabs}
\usepackage[numbers,sort&compress]{natbib}

\usepackage{physics}
\usepackage{amsthm}
\usepackage{amssymb}
\usepackage{dsfont}
\usepackage{amsmath}
\usepackage{lettrine}
\usepackage{bm}
\usepackage{balance}
\makeatletter
\theoremstyle{plain}

\theoremstyle{plain}

\theoremstyle{plain}
\newtheorem*{prop*}{\protect\propositionname}

\usepackage{braket}

\usepackage{tikz}
\usepackage{rotating}

\def\la{\lambda}

\def\la{\langle}
\def\ra{\rangle}

\newcommand{\beq}{\begin{equation}}
\newcommand{\eeq}{\end{equation}}
\newcommand{\beqa}{\begin{eqnarray}}
\newcommand{\eeqa}{\end{eqnarray}}

\makeatletter

\begin{document}
\title{Quantum Alchemy and Universal Orthogonality Catastrophe in One-Dimensional Anyons}

\author{Naim E. Mackel\href{https://orcid.org/0000-0001-7902-570X}{\includegraphics[scale=0.05]{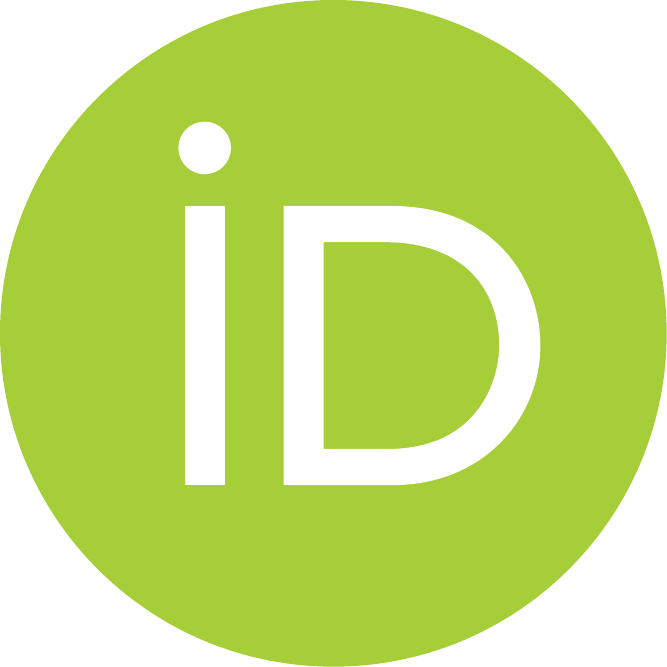}}}
\email{naimmackel@outlook.de}
\address{Department of Physics and Materials Science, University of Luxembourg,
L-1511 Luxembourg, Luxembourg}

\author{Jing Yang\href{https://orcid.org/0000-0002-3588-0832} {\includegraphics[scale=0.05]{orcidid}}}
\email{jingyangyzby@gmail.com}
\address{Department of Physics and Materials Science, University of Luxembourg,
L-1511 Luxembourg, Luxembourg}
\address{Nordita, KTH Royal Institute of Technology and Stockholm University, Hannes Alfvéns vag 12, 106 91 Stockholm, Sweden}

\author{Adolfo del Campo\href{https://orcid.org/0000-0003-2219-2851}{\includegraphics[scale=0.05]{orcidid}}}
\email{adolfo.delcampo@uni.lu}
\address{Department of Physics and Materials Science, University of Luxembourg,
L-1511 Luxembourg, Luxembourg}
\address{Donostia International Physics Center, E-20018 San Sebasti\'an, Spain}

\begin{abstract}
Many-particle quantum systems with intermediate anyonic exchange statistics are supported in one spatial dimension. 
In this context, the anyon-anyon mapping is recast as a continuous transformation that generates shifts of the statistical parameter $\kappa$.  We characterize the geometry of quantum states associated with different values of $\kappa$, i.e., different quantum statistics.  While states in the bosonic and fermionic subspaces are always orthogonal, overlaps between anyonic states are generally finite and exhibit a universal form of the orthogonality catastrophe governed by a fundamental statistical factor, independent of the microscopic Hamiltonian.  We characterize this decay using quantum speed limits on the flow of $\kappa$, illustrate our results with a model of hard-core anyons, and discuss possible experiments in quantum simulation.
\end{abstract}

\section{Introduction}

The spin-statistics theorem dictates that in the familiar three-dimensional world particles are either bosons or fermions. 
In lower spatial dimensions, however, intermediate exchange statistics are allowed, giving rise to the existence of anyons. 
Anyons are characterized by many-body wavefunctions that are not necessarily fully symmetric or antisymmetric, 
but can pick up an arbitrary phase factor under particle exchange. 

A model of two-dimensional Abelian anyons was first introduced by Leinaas and  Myrheim~\cite{Leinaas77} and further elaborated by Wilczek \cite{Wilczek82a,Wilczek82b}.
The study of two-dimensional anyons has grown into a substantial body of literature \cite{WilczekSaphere89,Khare05}. This anyonic behavior should not be confused with the notion of generalized exclusion statistics, as described by Haldane and Wu, possible in arbitrary spatial dimensions \cite{Haldane91,Wu94,MurthyShankar94}.
Decades later it was appreciated that intermediate exchange statistics is also possible in one spatial dimension \cite{Aglietti96,Kundu99}. 
Several models of interacting one-dimensional anyons have been characterized including contact interactions as well as hardcore \cite{Kundu99,Batchelor06,Girardeau06,Batchelor06PRB,delcampo08} and finite-range potentials.
While the inclusion of spin degrees of freedom is possible, we shall focus on spinless (or spin-polarized) quantum states of one-dimensional anyons.
The resulting families of anyons are labeled by the statistical parameter $\kappa$ that governs the statistical phase factor arising from particle exchange.
For $\kappa=0$ one recovers fully symmetric wavefunctions while the case $\kappa=\pi$ corresponds to antisymmetric fermionic wavefunctions. 
Proposals to realize models of one-dimensional anyons have been put forward using optical and resonator lattices as quantum simulators \cite{Keilmann11,Longhi12,GreschnerSantos15,Eckardt16,ZhangGreschner17,YuanFan17,Greschner18}. An experimental realization of one-dimensional anyons has been reported using ultracold atoms in an optical lattice \cite{Kwan23}.
In these scenarios, the statistical parameter $\kappa$ is not fixed and it is possible to conceive experiments in which its value is tuned dynamically. 
Such prospects pave the way for quantum alchemy, i.e., the transmutation of particles of one kind into another, such as bosons into anyons \cite{WilczekSaphere89}.

The fact that the permutation of particles in one spatial dimension is necessarily interwoven with interparticle interactions gives rise to the existence of several dualities, generalizing the celebrated Bose-Fermi mapping introduced by Girardeau between strongly interacting bosons and free fermions \cite{Girardeau60}. The description of one-dimensional hardcore anyons is possible using the anyon-anyon mapping, which relates states with different values of the statistical parameter $\kappa$ \cite{Girardeau06}. This generalized duality has spurred the investigation of hardcore anyons, making it possible to characterize efficiently ground-state correlations~\citep{Santachiara07,HaoZhangChen08,HaoZhangChen09,HaoChen12},
finite-temperature behavior~\citep{Patu08,Patu08b}, 
and their nonequilibrium dynamics~\citep{delcampo08,Piroli17,LiuGorshkov18}.

In this context, we associate the anyon-anyon mapping with a continuous transformation describing shifts of the statistical parameter. We show that under statistical transmutation, permutation symmetry yields a universal form of the orthogonality catastrophe governing the decay of quantum state overlaps in a way that is independent of the underlying system Hamiltonian. 
This universal behavior further determines the distinguishability of anyonic quantum states and the quantum geometry of the space of physical quantum states encompassing different quantum statistics.

\medskip

\section{Anyon-anyon mapping as a continuous transformation}
When the spin degrees of freedom can be ignored (e.g., in a fully polarized state), the spatial wavefunctions of bosons and fermions are respectively fully symmetric and antisymmetric with respect to particle exchange. No permutation-symmetric operator can couple them and thus exchange statistics imposes a superselection rule in which the Hilbert space of a physical system of identical particles is the direct sum of the bosonic and fermionic subspaces. However, the importance of mappings between different sectors has been long recognized in many-body physics. The celebrated Bose-Fermi duality provides a prominent example, relating wavefunctions of hard-core bosons $\Psi_{\rm HCB}$ to that of spin-polarized fermions $\Psi_{\rm F}$ in one spatial dimension: $\psi_{\rm HCB}=\prod_{i<j}{\rm sgn}(x_{ij})\Psi_{\rm F}$ where $x_{ij}=x_i-x_j$. 
The extension of the Bose-Fermi mapping \cite{Girardeau60} to anyons was put forward by Girardeau \cite{Girardeau06}, building on earlier results by Kundu \cite{Kundu99} and applied to the construction of anyonic wavefunctions from either bosonic or fermionic states. For instance, given $\Psi_{\rm F}$ one obtains the corresponding state of hard-core anyons $\Psi_{ \kappa}$ with statistical parameter $\kappa$ using the mapping $\Psi_{\kappa}=\exp(-{\rm i} \frac{\kappa}{2}\sum_{i<j}{\rm sgn}(x_{ij}))\Psi_{\rm HCB}$ \cite{Girardeau06}. Although models with softcore interactions are possible (as in the case of the well-studied Lieb-Liniger anyons), the hardcore condition by which $\Psi_{\kappa}=0$ when $x_{ij}=0$ is rather ubiquitous. It arises when the strength of contact interactions is divergent (i.e., as a limit of Lieb-Liniger anyons with repulsive interactions), for pairwise power-law potentials (e.g., $V= \sum_{i<j}\lambda/|x_{ij}|^\alpha$ with $\lambda,\alpha>0$, as in the case of  Calogero-Sutherland anyons with $\alpha=2$ \cite{Girardeau06}), and with other interaction potentials satisfying $V\rightarrow+\infty$ as $x_{ij}\rightarrow 0$.

Here, we consider a natural generalization transforming anyonic wavefunctions $\Psi_{\kappa'}$  with statistical parameter $\kappa'$  into anyonic wavefunctions $\Psi_{\kappa}$ with statistical parameter $\kappa$ via the linear mapping 
\beqa
\Psi_{\kappa}=\hat{\mathcal{A}}(\kappa,\kappa')\Psi_{\kappa'}. 
\eeqa
In doing so, the anyon-anyon mapping $\hat{\mathcal{A}}(\kappa,\kappa')$   is associated with a continuous unitary transformation in which the generator
\beqa
\hat{G}=\frac{1}{2}\sum_{i<j}{\rm sgn}(x_{ij}),
\eeqa
induces shifts of the statistical parameter $\kappa$. As $\hat{G}$ does not depend on it explicitly, the mapping is given by the unitary, $\hat{\mathcal{A}}(\kappa,\kappa')=\hat{\mathcal{A}}(\kappa-\kappa')=\exp\left[-{\rm i}(\kappa-\kappa')\hat{G}\right]$, 
and thus satisfies all the group properties, including the existence of the identity $ \hat{\mathcal{A}}(0)=\mathbb{I}$,  inverse $\hat{\mathcal{A}}(\kappa)^{-1}=\hat{\mathcal{A}}(-\kappa)=\hat{\mathcal{A}}(\kappa)^\dag$, and group multiplication $\hat{\mathcal{A}}(\kappa)\hat{\mathcal{A}}(\kappa')=\hat{\mathcal{A}}(\kappa+\kappa')$, when $\kappa$ takes values on the real line. We note however that it suffices to consider the domain $\kappa\in[0,2\pi)$ upon identifying $\kappa+2n\pi\sim\kappa$ for any integer $n\in\mathbb{Z}$. 

\medskip 

\section{Many-anyon state overlaps}
Consider $N$ one-dimensional spinless hardcore anyons in an arbitrary quantum state $\Psi_\kappa$ belonging to the Hilbert space $\mathcal{H}_\kappa$, which is a subspace with anyonic exchange symmetry of the Hilbert space of square-integrable functions $\mathcal{L}^2(\mathbb{R}^N)$. Specifically, $\Psi_\kappa(x_1,\dots,x_i,x_{i+1},\dots,x_N)=e^{{\rm i} \kappa{\rm sgn}(x_i-x_{i+1})}\Psi_\kappa(x_1,\dots,x_{i+1},x_i,\dots,x_N)$.
The application of the  mapping $\hat{\mathcal{A}}(\delta)$ on  an anyonic state $\ket{\Phi_\kappa}$ introduces a unitary flow of the state, leading to a distinguishable state $|\Phi_{\kappa+\delta}\ra \equiv e^{-{\rm i} \hat{G} \delta }|\Phi_{\kappa}\ra$ with  the statistical parameter shifted by $\delta$.  Anyons with statistical parameter $\kappa$   are thus transmuted into anyons with statistical parameter $\kappa'$, motivating the term ``quantum alchemy'' for such transformation. Note that whenever $\kappa+\delta=\pi$, $|\Phi_{\kappa+\delta}\ra$ describes a fermionic wave function, which vanishes at the contact points where at least two coordinates coincide. Thus, $|\Phi_{\kappa}\ra$ must vanish at the contact points, obeying a hard-core constraint. 
We note that the family of hard-core anyons is not restricted to contact interactions but can accommodate, e.g., power-law interactions \cite{Girardeau06}.

Having justified the flow of the states in the Hilbert space of identical particles, we aim at characterizing the quantum geometry of state space and ask what is the distance between the state in $\mathcal{H}_{\kappa+\delta}$ and the original state in $\mathcal{H}_{\kappa}$.  To this end, we consider $\Psi_\kappa \in \mathcal{H}_\kappa$ and compute the survival amplitude defined by the overlap
\beqa
\la\Psi_\kappa|\Phi_{\kappa+\delta}\ra=\la\Psi_\kappa|e^{-{\rm i} \hat{G} \delta }|\Phi_{\kappa}\ra.
\eeqa 
On a given sector $\mathcal{R} : x_{\mathcal{R}(1)}>x_{\mathcal{R}(2)}> \dots > x_{\mathcal{R}(N)}$, the action of the anyon-anyon mapping can be replaced by a phase factor $\omega_\delta(\mathcal{R} )$.
We thus consider a generalized Heaviside step function $\mathds{1}_{\mathcal{R}}$:
\begin{widetext}
\beqa \label{eq:1func}
\mathds{1}_\mathcal{R} = \mathds{1}_{x_{\mathcal{R}(1)}>x_{\mathcal{R}(2)}> \dots > x_{\mathcal{R}(N)}} \equiv \begin{cases} 1 & \textrm{if } x_{\mathcal{R}(1)}>x_{\mathcal{R}(2)}> \dots > x_{\mathcal{R}(N)} \\ 0 & \textrm{otherwise} \end{cases}.
\eeqa
Making use of it, we note that survival amplitude can be written as the sum over $N!$ sectors, associated with permutations over the symmetric group $S_N$ (see Appendix ~\ref{I-N} for more details),
\beqa
\la\Psi_\kappa|\Phi_{\kappa+\delta}\ra=\sum_{\mathcal{R} \in S_N} \omega_\delta(\mathcal{R}) I_\kappa(\mathcal{R}),
\eeqa 
where 
\begin{equation}
\omega_\delta(\mathcal{R})\equiv e^{-\text{i}\frac{\delta}{2}\sum_{i<j}\text{sgn}\left(x_{i}-x_{j}\right)}\big |_{x_{\mathcal{R}(1)}>x_{\mathcal{R}(2)}> \dots > x_{\mathcal{R}(N)}},\label{eq:omg-def}
\end{equation}
and $I_\kappa(\mathcal{R})$ involves the $N$-dimensional integral
\beqa
I_\kappa(\mathcal{R})=\int_{\mathbb{R}^N} \prod_{i=1}^N \dd{x_i} \Psi^{*}_\kappa(x_1, \cdots x_N) \Phi_\kappa(x_1, \cdots x_N) \mathds{1}_\mathcal{R}.
\eeqa
\end{widetext}
Its explicit evaluation is challenging as it involves  $N!$ $N$-dimensional integrals. However, an evaluation in closed form is possible making use of permutation symmetry. 
We note that for arbitrary pairs of $\Psi_\kappa, \Phi_\kappa$ and $\Psi_{\kappa'}, \Phi_{\kappa'}$ connected by the anyon-anyon mapping, the corresponding integrals are equal,  i.e., $\forall \kappa,\kappa' \in \mathbb{R}: I_\kappa(\mathcal{R})=I_{\kappa'}(\mathcal{R})\equiv I(\mathcal{R})$. In addition, integrals evaluated at  different sectors are all equal, i.e., $\forall \mathcal{R},\mathcal{R}' \in S_N: I(\mathcal{R})=I(\mathcal{R}')$, thanks to the permutation symmetry of the integrand $ \Psi^{*}_\kappa(x_1, \cdots x_N) \Phi_\kappa(x_1, \cdots x_N)$. 
From the resolution of the identity $\sum_{\mathcal{R} \in S_N} \mathds{1}_\mathcal{R}=\mathbb{I}_{\mathbb{R}^N}$, we conclude that $I(\mathcal{R})=\la \Psi_\kappa | \Phi_\kappa \ra/ N!$, and thus, 
\beqa
\la\Psi_\kappa|\Phi_{\kappa+\delta}\ra=\la \Psi_\kappa | \Phi_\kappa \ra \frac{1}{N!}\sum_{\mathcal{R} \in S_N} \omega_\delta(\mathcal{R}).
\eeqa 
%
%
\begin{figure}[t]
\centering
\includegraphics[width=0.8\linewidth]{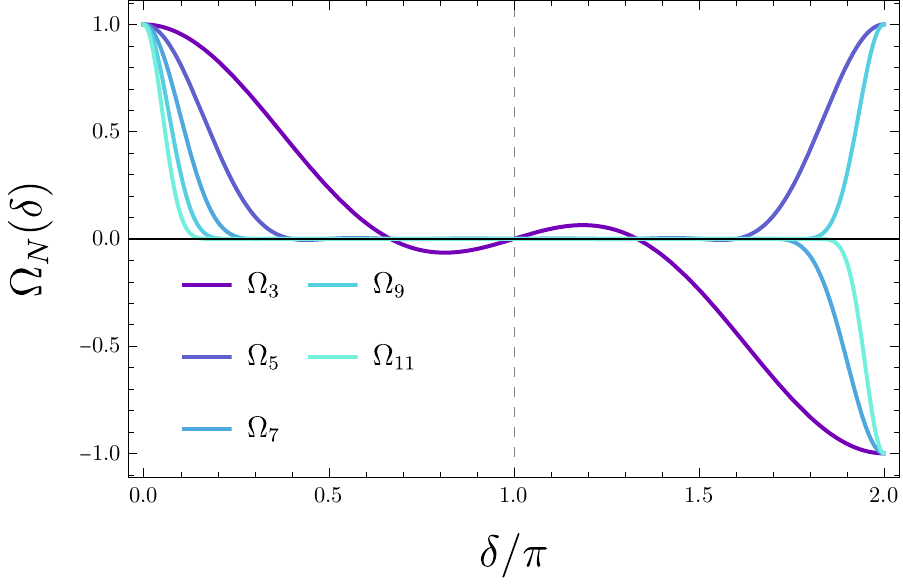}
\vspace{-0.5cm}
\caption{{\bf Statistical contribution to the survival amplitude at different system sizes.} As the system size $N$ increases, the statistical contribution $\Omega_N$ to the overlap far from the $\delta = 0, 2 \pi$ decays quickly, and becomes progressively steeper within this neighborhood. Note that $\delta\in[0,2\pi]$ is the relevant part to plot, since $\Omega_N$ is an even function and $4 \pi$-periodic.}
\label{Fig1Omega}
\end{figure}
%
The overlap between anyonic states with different quantum statistics depends on the state overlap   $\la \Psi_\kappa | \Phi_\kappa \ra$ between states with common $\kappa$ and on an additional contribution from the shift $\delta$ of the statistical parameter
\beqa
\Omega_N(\delta) := \frac{1}{N!}\sum_{\mathcal{R} \in S_N} \omega_\delta(\mathcal{R}).
\eeqa 
As shown in Appendix~\ref{App:4}, this yields the recursion relation
\beqa
\Omega_N(\delta) & = \frac{1}{N} \sum_{n=0}^{N-1} e^{-{\rm i} \frac{\delta}{2} (N-1-2n)} \Omega_{N-1}(\delta),
\eeqa 
making it possible to find  by iteration the exact close-form expression 
\begin{equation} \label{eq:OmegaP}
\Omega_N(\delta) = \frac{1}{N!} \prod_{n=2}^{N} \frac{\sin\left( \frac{n \delta}{2} \right)}{\sin\left( \frac{\delta}{2} \right)},
\end{equation}
shown in Fig. \ref{Fig1Omega} as a function of the statistical shift for different values of $N$.
Under the sole consideration of a statistical shift $\delta$, i.e., choosing $\ket{\Phi_\kappa} = \ket{\Psi_\kappa}$, the survival amplitude of an initial state under the flow induced by the anyon-anyon mapping collapses to $\Omega_N(\delta)$. This is a key fundamental result from which our subsequent analysis follows.

\section{Quantum speed limits on the flow of the statistical parameter}
Quantum speed limits (QSLs) provide lower bounds on the time required for a process to unfold.  While introduced in quantum dynamics \cite{Mandelstam45,Margolus98},  QSLs apply to the flow of quantum states under other continuous transformations described as a one-parameter flow \cite{Braunstein96,Margolus21}.  Given that the anyon-anyon mapping can be described by a unitary flow, the decay of the quantum state overlap under shifts of $\kappa$ is subject to generalizations of QSLs. In quantum dynamics, the Mandelstam-Tamm QSL and the Margolus-Levitin QSL bound the minimum time for the evolving state to become orthogonal to the initial state in terms of the energy dispersion and the mean energy of the initial state, respectively \cite{Mandelstam45,Margolus98,Deffner17}.   They can be generalized in the current context as shown in Appendix~\ref{QSL}, to identify a bound on the minimum $\kappa$-shift required for the initial state $\ket{\Psi_{\kappa}}$ to be transmuted into an orthogonal state. Specifically, the generalized QSLs  take the form
\beqa \label{eq:QSL}
\kappa\geq \kappa_{\rm MT}&=&\frac{\pi}{2 \sqrt{\langle \hat{G}^2 \rangle - \langle \hat{G} \rangle^2}},\\
\kappa\geq \kappa_{\rm ML}&=& \frac{\pi}{2 (\langle \hat{G} \rangle - G_0)},
\eeqa
where $G_0=- \frac{N(N-1)}{4}$ is the lowest eigenvalue of $\hat{G}$. \textcolor{black}{At variance with the familiar case concerning time evolution under a given Hamiltonian, the generator $\hat{G}$ is uniquely set by the anyon-anyon mapping: there is no freedom in its choice.}
The brackets in $\la \hat{G}^n\ra$ can be used to denote the expectation value over the initial state $\ket{\Psi_{\kappa}}$,  or equivalently, the average over the $N!$  sectors in configuration space, given that the value of $\la \hat{G}^n\ra$ is the same for any wavefunction of indistinguishable particles.      
As a result, and at variance with the conventional QSL, the characteristic shifts of the statistical parameter are independent of the quantum state, and are universal, being solely governed by permutation symmetry.
The universal factor $\Omega_N(\delta)$ can be viewed as the generating function of the moments of the generator $G$ over the initial state $\ket{\Psi_{\kappa}}$, i.e,
$ \la \hat{G}^n\ra=i^n\frac{d^n\Omega_{N}(\delta)}{d\delta^n }|_{\delta=0}$,
from which one  obtains $\la \hat{G}\ra=0$, $\la \hat{G}^2\ra=- \frac{N (2N^2+3N-5)}{72}$.
Thus, we find the universal lower bounds
\beqa
\kappa\geq \kappa_{\rm MT} &=& \frac{3 \sqrt{2} \pi}{\sqrt{N (2N^2+3N-5)}},\\
\kappa\geq\kappa_{\rm ML} &=&  \frac{2 \pi}{N(N-1)}.
\eeqa
For unitary flows generated by time-independent generators and pure initial states, the 
 MT speed limit can be expressed as $\kappa_{\rm MT}=\pi/\sqrt{F^Q(\kappa)}$, where $F^Q(\kappa)\equiv4(\langle \hat{G}^2 \rangle - \langle \hat{G} \rangle^2)$ is known as the quantum Fisher information, which describes the Riemannian geometry of the quantum state space \cite{braunstein1994}. Specifically, it is tied to the Fubini-Study metric on the state manifold parameterized by $\kappa$, i.e.,  $ds^2=\frac{1}{4}F^Q(\kappa)d\kappa^2$, where $ds$ is the infinitesimal distance between $\ket{\Psi_{\kappa}}$ and $\ket{\Psi_{\kappa+d\kappa}}$. The quantum Fisher information also characterizes the variance in practical estimation theory via the quantum Cram\'er-Rao bound~\cite{paris2009}. For generators with linear interactions between particles, it has been shown that the quantum Fisher information scales at most as $(N\ln N)^2$~\cite{boixo2007,yang2021}, where $N$ is the number of particles. In our case, $F^Q(\kappa)$ scales as $N^3$, reminiscent of the super-Heisenberg scaling in the context of parameter estimation quantum metrology with a nonlinear generator $\hat{G}$ \cite{boixo2007}.
 \textcolor{black}{ In addition,   Anandan and Aharanov showed that the saturation of the MT bound occurs when the state evolves along a shortest Fubini-Study geodesic \cite{AnandanAharonov90}.
The fact that MT QSL is not saturated indicates that the unitary flow induced by the anyon-anyon mapping does not trace a shortest geodesic in the space of physical states. This raises the question as to whether such flow is optimal in any certain sense. }

\begin{figure}[t]
\centering
\includegraphics[width=0.8\linewidth]{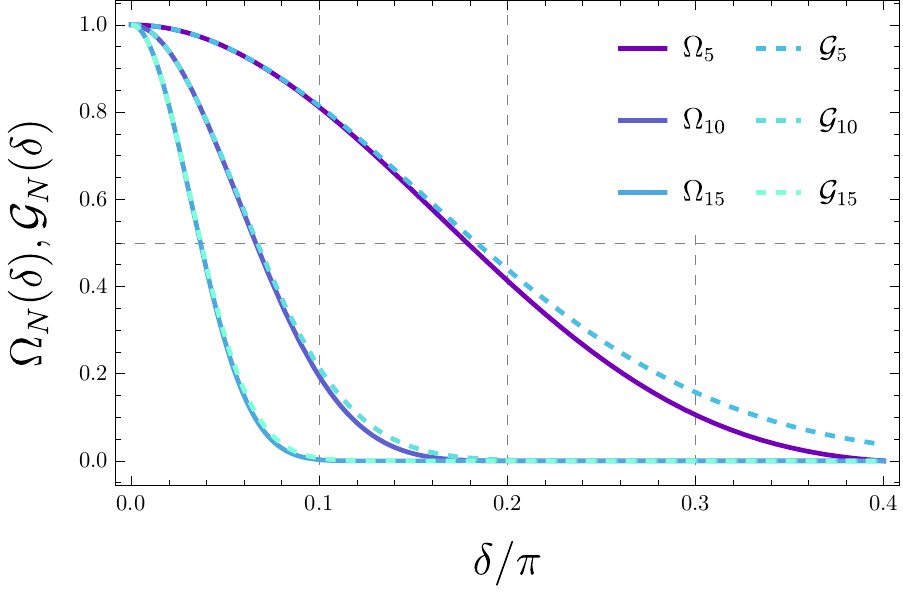}
\vspace{-0.5cm}
\caption{{\bf Comparison between the exact statistical contribution and its Gaussian approximation.} As the system size $N$ increases, $\Omega_N$ is increasingly more precisely approximated by $\mathcal{G}_N$.}
\label{Fig2QSL}
\end{figure}

\medskip

\section{Universal Orthogonality Catastrophe}
Many-body eigenstates are highly sensitive to local perturbations. For fermions, this dependence is extensive in the system size and is known as the orthogonality catastrophe  \cite{Anderson67}. This phenomenon has been analyzed in the case of particles obeying generalized exclusion statistics 
\cite{delCampo16,Ares18}. Its occurrence has been further related to QSL in \cite{Fogarty20}. 
It is natural to explore analogs of it under the transmutation of particles. For the case at hand, we note that hard-core anyons are related to the one-dimensional spin-polarized Fermi gas by the anyon-anyon mapping.  Further, the generator $ \hat{G}$ of $\kappa$ shifts is two-body and spatially nonlocal. 
For an infinitesimal flow of the statistical parameter, we show in Appendix~\ref{QSL} that the overlap between anyonic states decays as
\begin{equation} \label{eq:Gauss}
\Omega_N(\delta) \approx \exp\left[-\frac{\delta^2}{2} \frac{N (2N^2+3N-5)}{72}\right]=\mathcal{G}_N(\delta).
\end{equation}
A comparison between this Gaussian approximation and the exact result (\ref{eq:OmegaP}) is shown in Fig \ref{Fig2QSL}. 
This overlap decays rapidly as the particle number $N$ is increased. Given that the variance $\sigma^2=\la \hat{G}^2\ra$ governs this decay, one may be tempted to conclude that the MT bound governs the orthogonality catastrophe, as proposed in \cite{Fogarty20}.
Yet,  from the explicit expression  in Eq. (\ref{eq:OmegaP}), it is found that  $\Omega_N(\delta) $ vanishes identically for $\delta\in\mathcal{Z}_N$, where
 \begin{equation} \label{eq:zeroset}
\mathcal{Z}_N = \left\{ \frac{2\pi k}{n} \Big| \, n = 2,\dots,N; \; k = 1,\dots,n-1 \right\}.
\end{equation} 
Note that the values $\delta=0,2\pi$ are not zeroes and are thus excluded.
For a given $N$, the interval where these zeros accumulate is given by
$I = \left[ \frac{2 \pi}{N}, 2 \pi - \frac{2 \pi}{N} \right]$.
In the thermodynamic limit $N \to \infty$, the values on this interval become all zero, in addition the interval becomes $I_{N \to \infty} = (0, 2 \pi)$, while the values in ${0, 2 \pi}$ remain unchanged. Hence, one finds
\begin{equation} \label{eq:Oinfty}
|\Omega_{N \to \infty}| = \delta_{\delta, 2 \pi k} \textrm{  , where } k \in \mathbb{Z}.
\end{equation} 
\begin{figure}[t]
\centering
\includegraphics[width=0.8\linewidth]{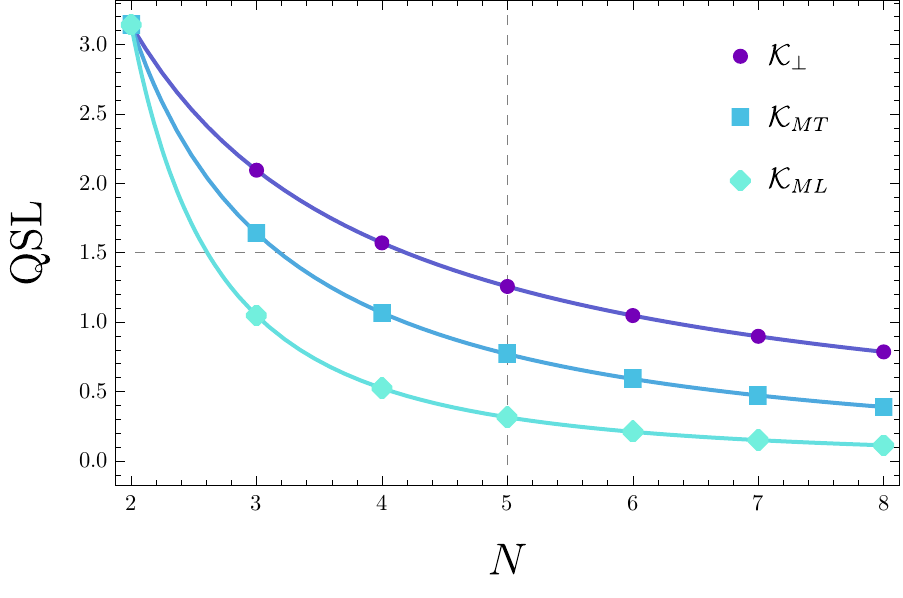}
\vspace{-0.5cm}
\caption{{\bf Comparison of the two QSL estimates with the exact determined QSL.} Minimum shift of the statistical parameter $\kappa$  estimated by the generalized QSL in comparison with the exact values determined as zeroes of the overlap $\Omega_{N}(\delta)$ between anyonic many-body states.  The dependence with the system size $N$ of the orthogonalization $\kappa$-shift is incorrectly predicted by the   MT and ML bounds, which are too conservative and never saturated in the flow of particle transmutation.}
\label{Fig3QSL}
\end{figure}

The first zero of  $\Omega_{N}(\delta)$ for a given $N$ describes the exact $\kappa$-shift for the anyonic state to be transmuted to
an orthogonal state and is given by
\begin{equation}
\label{exactkshift}
\kappa_{\perp}\equiv\frac{2\pi}{N}. 
\end{equation}
Despite the inverse scaling with the number of particles, neither the MT nor the ML bound predicts the correct scaling (\ref{exactkshift}). Direct comparison between the QSLs and $\kappa_{\perp}$ is displayed in Fig.~\ref{Fig3QSL}. As illustrated, for any $N>2$, the chain of inequalities  $\kappa_{\rm ML} <\kappa_{\rm MT}<\kappa_{\perp}$ is fulfilled and therefore $\kappa_{\rm MT} $ is tighter than the ML bound. And yet, the dependence of the minimum $\kappa$ shift required for orthogonality is incorrectly estimated by QSL, indicating the need for caution in using QSL as a proxy for orthogonality catastrophe  \cite{Fogarty20}.

\textcolor{black}{Our analysis of the orthogonality catastrophe shed new light on the nature of superselection rules in one-dimensional hardcore anyons. While coherent quantum superpositions between states of bosons and fermions are forbidden (and more generally, when $\Omega_N(\pi)=0$) superpositions between anyons with different statistical parameters satisfying $\delta\neq \pi$ are generally possible at finite $N$. Only in the limit $N\rightarrow\infty$, do such superpositions cease to occur. In this setting, superselection rules are derived and emerge from quantum information geometry and the anyon-anyon mapping. It would be desirable to generalize this approach to higher dimensions. However, this faces the well-known difficulty of extending Bose-Fermi and anyon-anyon dualities when the notion of particle ordering no longer holds.}

\section{Single-qubit interferometry}
Quantum simulators exploit a physical platform to mimic the behavior of a system of interest \cite{Georgescu14}. While the physical platform is governed by the laws of physics, the simulated system can explore alternative laws and operations unphysical at the platform level \cite{Casanova11}, making possible the simulation of quantum alchemy. 
 Given an experimental setup for the quantum simulation of hard-core anyons \cite{Keilmann11,Longhi12,GreschnerSantos15,Eckardt16,ZhangGreschner17,YuanFan17,Greschner18},  we next consider an experimental protocol implementing their statistical transmutation to determine the distinguishability of anyonic quantum states. The protocol measures the overlap $\la\Psi_\kappa|\Psi_{\kappa+\delta}\ra$, which becomes $\Omega_N(\delta)$, provided that $|\Psi_\kappa \ra$ is normalized. 
 It has been argued that the nature of anyonic statistics in one dimension is dynamical and not topological \cite{Harshman22}. Proposals to simulate 1D anyons exploit this feature \cite{Keilmann11,Longhi12,GreschnerSantos15,Eckardt16,ZhangGreschner17,YuanFan17,Greschner18}. To estimate the overlap between wavefunctions one can thus consider making use of an ancilla, providing an auxiliary degree of freedom for control, as recently done to probe QSL in the laboratory \cite{Alberti21}. A simpler protocol is that of single-qubit interferometry as described to measure the characteristic function of many-body observables \cite{Xu19}. 
 %
\begin{figure}[t]
\centering
\includegraphics[width=0.8\linewidth]{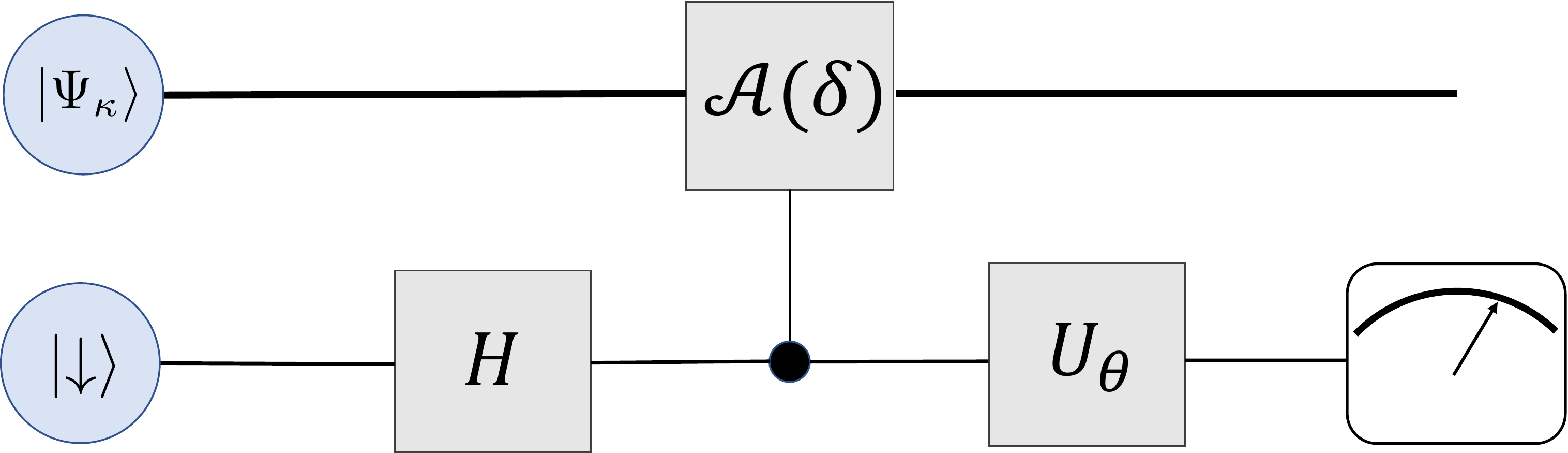}
\centering
\caption{{\bf Experimental protocol of the single-qubit interferometry.} The protocol measures the overlap of two many-body anyonic wave functions $|\Psi_{\kappa}\rangle$ and $|\Psi_{\kappa+\delta}\rangle \equiv \mathcal{A}(\delta)|\Psi_{\kappa'}\rangle$.  An ancilla qubit is initially prepared in the ground state $\ket{\downarrow}$. After the Hadamard gate, a controlled-$\mathcal{A}(\delta)$ gate is implemented so that the state becomes $(|\Psi_{\kappa+\delta}\ra\otimes|\uparrow\ra+|\Psi_{\kappa}\ra\otimes|\downarrow\ra)/\sqrt{2}$. Taking the single-qubit rotation $U_{\theta}$ to be $U_{\theta}=e^{{\rm i} \sigma_y \pi/4}$ and $U_{\theta}=e^{{\rm i} \sigma_x \pi/4}$ and then measuring in the computational basis $\{ \ket{\uparrow},\, \ket{\downarrow}\}$ at the end, one can obtain the real and imaginary part of $\langle \Psi_{\kappa}\big| \Psi_{\kappa+\delta}\rangle$, respectively.}
\label{Fig4SQI}
\end{figure}
%
%
 Single-qubit interferometry provides the means to measure the overlap between many-body anyonic wave functions characterized by different statistical parameters when connected by the anyon-anyon mapping, and thus to determine the universal factor $\Omega_N(\delta)$. The procedure is illustrated in Fig.~\ref{Fig4SQI}.  An ancilla qubit is initially prepared in the ground state (eigenstate of $\sigma_z$ with eigenvalue $-1$) and is brought to a state with equal-weight coherent quantum superposition through the Hadamard gate $H$. Then, it is coupled to the many-body anyonic state through the unitary flow conditioned on the ancilla state, i.e.,  the controlled-$\mathcal{A}(\delta)$ gate: $\mathcal{A}(\delta)\otimes|\uparrow\ra \la \uparrow|+\mathbb{I}\otimes|\downarrow\ra \la \downarrow|$.  After this step, the composite state becomes 
 $(|\Psi_{\kappa+\delta}\ra\otimes|\uparrow\ra+|\Psi_{\kappa}\ra\otimes|\downarrow\ra)/\sqrt{2}$. 
 Finally, $\sigma_x$-measurements can be performed, yielding 
 \begin{equation}
 P(+x)=\frac{1}{2} [1+{\rm Re}\langle \Psi_{\kappa}\big| \Psi_{\kappa+\delta}\rangle]=\frac{1}{2} [1+\Omega_N(\delta)]
 \end{equation} 
 and $P(-x)=1-P(+x)$. The universal factor $\Omega_N(\delta)$ can be inferred by measuring the interference fringes as a function of the statistical parameter $\delta$.  As discussed previously,  $\Omega_N(\delta)$ vanishes identically when $N \to \infty$ due to the orthogonality catastrophe, and therefore the interference pattern will be suppressed for $\sigma_x$-measurements in the thermodynamic limit. Similarly, if $\sigma_y$-measurements are performed, $P(+y)$ is related to  the imaginary part of the overlap $\langle \Psi_{\kappa}\big| \Psi_{\kappa+\delta}\rangle$, which identically zero in this case. No interference pattern is observed in $\sigma_y$ measurements, regardless of the number of particles.
Practically, both $\sigma_x$ and $\sigma_y$ measurements can be implemented by $\pi/2$ pulses defined as $U_{\theta}=e^{{\rm i} \sigma_y \pi/4}$ and $U_{\theta}=e^{{\rm i} \sigma_x \pi/4}$ respectively,  followed by a measurement in the computational basis $\{ \ket{\uparrow},\, \ket{\downarrow}\}$, as shown in Fig.~\ref{Fig4SQI}.    

\section{Conclusion}
We have described the transmutation of hardcore anyons associated with the continuous unitary flow of the statistical parameter $\kappa$.
While the space of physical states of bosons and fermions is governed by well-known superselection rules,  we show that anyonic states are generally not orthogonal and are only partially distinguishable. 
Many particle states with different values of the statistical parameter exhibit a universal form of the orthogonality catastrophe generated by the two-body operator entering the anyon-anyon mapping. The decay overlap is underestimated by generalizations of the Mandelstam-Tamm and Margolus-Levitin quantum speed limits, which are too conservative in this context.
 Indeed, the overlap decay is universal, independent of the system Hamiltonian or the specific quantum state, and is solely governed by a fundamental statistical factor. 
 Building on the schemes proposed for the quantum simulation of hard-core anyons with a tunable statistical parameter $\kappa$, the universal form of the orthogonality catastrophe and the quantum state geometry associated with it can be probed by making use of single-qubit interferometry and related schemes recently implemented for the experimental study of quantum speed limits.

\section*{Acknowledgements}
The authors are indebted to Federico Balducci, L\'eonce Dupays,  \'I\~nigo L. Egusquiza, and Federico Roccati for comments on the manuscript.

\onecolumn
\appendix
\part*{Appendix}
\section{Spectral properties of the generator $\hat{G}$}
Consider the Hilbert space of $N$ spinless indistinguishable particles in one spatial dimension $\mathcal{H}=\bigotimes_{i=1}^N\mathcal{L}^2(\mathbb{R},dx_i)/S_N\cong\mathcal{L}^2(\mathbb{R}^N/S_N)$.
We note the resolution of the identity
\beqa
\sum_{\mathcal{R} \in S_N} \mathds{1}_\mathcal{R}=\mathds{1}_{\mathbb{R}^N}.
\eeqa
that splits the configuration space into $N!$ sectors.
Each sector is an eigenspace of the linear self-adjoint operator
\beqa
\hat{G}=\frac{1}{2}\sum_{i<j}{\rm sgn}(x_{ij}),
 \eeqa
satisfying
\beqa
\hat{G}\, \mathds{1}_\mathcal{R}=g_\mathcal{R}\,\mathds{1}_\mathcal{R}.
\eeqa
We further note that  
\beqa
\mathds{1}_\mathcal{R}\,\mathds{1}_\mathcal{R'}= \mathds{1}_\mathcal{R}\delta_{\mathcal{R}\,\mathcal{R}'}.
\eeqa
As a result, the generator $\hat{G}$ admits the spectral decomposition
\beqa
\hat{G}=\sum_{\mathcal{R} \in S_N} g_\mathcal{R}\,\mathds{1}_\mathcal{R}.
\eeqa
The pointwise spectrum of $\hat{G}$ reads
\beqa \label{eq:SpecG}
{\it Sp}(\hat{G})=\left\{-\frac{N(N-1)}{4}+n\, \bigg|\, n=\,0,1,\dots,\frac{N(N-1)}{2}\right\}.
\eeqa

\section{Unitary flow of the statistical parameter}
The generator $\hat{G}$ is independent of the statistical parameter $\kappa$. 
As a result, the anyon-anyon mapping requires no path ordering and is dependent only on the amplitude of the shift 
 \beqa
 \hat{\mathcal{A}}(\kappa,\kappa_0)=e^{-{\rm  i}(\kappa -\kappa_0)\hat{G}}= \hat{\mathcal{A}}(\kappa-\kappa_0),
 \label{Adef}
 \eeqa
 where $\kappa\in\mathbb{R}$.
 It reduces to the identity when $\kappa=\kappa_0$
 \beqa
  \hat{\mathcal{A}}(\kappa_0,\kappa_0)=  \hat{\mathcal{A}}(0)=\mathds{1}_{\mathbb{R}^N}
 \eeqa
and satisfies the composition (group multiplication) property
 \beqa
  \hat{\mathcal{A}}(\kappa_2,\kappa_0)=\hat{\mathcal{A}}(\kappa_2,\kappa_1)\hat{\mathcal{A}}(\kappa_1,\kappa_0).
 \eeqa
 Furthermore $\hat{\mathcal{A}}(\kappa_0,\kappa_1)\hat{\mathcal{A}}(\kappa_1,\kappa_0)=\mathbb{I}_{\mathbb{R}^N}$ and thus
 \beqa \label{eq:AdjointInverse}
 \hat{\mathcal{A}}(\kappa)^\dag= \hat{\mathcal{A}}(-\kappa)= \hat{\mathcal{A}}(\kappa)^{-1}.
 \eeqa
 $\hat{\mathcal{A}}(\kappa)$ is the analog of the time-evolution operator with a time-independent Hamiltonian in which the role of time is replaced by the statistical parameter 
 and the Hamiltonian is replaced by $\hat{G}$.
 By Stone's theorem, the one-parameter group $\mathcal{A}(\kappa)$ must be of the form explicit in its definition (\ref{Adef}) and satisfy the Schr\"odinger equation
 \beqa
 {\rm  i}\frac{d}{d\kappa} \hat{\mathcal{A}}(\kappa,\kappa_0)= \hat{G}\hat{\mathcal{A}}(\kappa,\kappa_0).
 \eeqa
 For completeness, we note the analog of the Heisenberg equation for an observable $\hat{O}$ under the flow of the statistical parameter $\kappa$.
 Given a quantum state $\Psi_\kappa= \hat{\mathcal{A}}(\kappa)\Psi_0$, we define the Heisenberg picture from the identity
 \beqa
 \la \Psi_\kappa|\hat{O}(0)|\Psi_\kappa\ra= \la \Psi_0|\hat{\mathcal{A}}(\kappa)^\dag\hat{O}(0)\hat{\mathcal{A}}(\kappa)|\Psi_0\ra=: \la \Psi_0|\hat{O}_H(\kappa)|\Psi_0\ra,
 \eeqa
i.e., $\hat{O}_H(\kappa)=\hat{\mathcal{A}}(\kappa)^\dag\hat{O}(0)\hat{\mathcal{A}}(\kappa)$.
 The derivation of (the analog of) the Heisenberg equation reads
  \beqa
 \frac{d}{d\kappa}\hat{O}_H(\kappa)=-{\rm  i}[\hat{O}_H(\kappa),\hat{G}].
 \eeqa
 given that $\hat{G}$ is independent of $\kappa$.

\section{Properties of the $N$-dimensional sector integrals \label{I-N}} 

\subsection{Independence from the statistical parameter}

Consider the integral
\beqa \label{eq:SecInt}
I_\kappa(\mathcal{R})=\int_{\mathbb{R}^N} \prod_{i=1}^N \dd{x_i} \Psi_\kappa^* \Phi_\kappa \mathds{1}_\mathcal{R},
\eeqa
through the anyon-anyon mapping a state with statistical parameter $\kappa$ can be rewritten as a function of any other parameter $\kappa'$
\beqa
\Psi_\kappa = \hat{\mathcal{A}}(\kappa',\kappa) \Psi_{\kappa'}.
\eeqa
This yields, for the integral (\ref{eq:SecInt}),
\beqa
I_\kappa(\mathcal{R})=\int_{\mathbb{R}^N} \prod_{i=1}^N \dd{x_i} \Psi_{\kappa'}^* \hat{\mathcal{A}}(\kappa',\kappa)^\dagger \hat{\mathcal{A}}(\kappa',\kappa) \Phi_{\kappa'} \mathds{1}_\mathcal{R} \stackrel{(\ref{eq:AdjointInverse})}{=} \int_{\mathbb{R}^N} \prod_{i=1}^N \dd{x_i} \Psi_{\kappa'}^* \Phi_{\kappa'} \mathds{1}_\mathcal{R} =I_{\kappa'}(\mathcal{R}).
\eeqa
Hence, the integral is independent from the statistical parameter,  $\forall \kappa,\kappa' \in \mathbb{R}: I_\kappa(\mathcal{R})=I_{\kappa'}(\mathcal{R})=I(\mathcal{R})$, as we may omit the specification of the parameter in the notation.

\subsection{Independence from the selected sector}

Consider the permutation operator $\hat{\mathcal{P}}$ defined by its action on an arbitrary $N$-variable function $f(x_1,x_2,\dots,x_3)$, i.e., 
\beqa
\hat{\mathcal{P}} f(x_1,x_2,\dots,x_N) = f(x_{\mathcal{P}(1)},x_{\mathcal{P}(2)},\dots,x_{\mathcal{P}(N)}),
\eeqa
where $\mathcal{P}$ is an arbitrary permutation of $\{1,2,\dots,N\}$. Within the integral 
\beqa
I(\mathcal{R})=\int_{\mathbb{R}^N} \prod_{i=1}^N \dd{x_i} \Psi_\kappa^* \Phi_\kappa \mathds{1}_\mathcal{R}
\eeqa
all variables are dummy. Furthermore, the integration over any of the $N$ variables is over the same domain $\mathbb{R}$. The integral remains unchanged under any permutation of variables
\beqa \label{eq:PermInt}
I(\mathcal{R})=\int_{\mathbb{R}^N} \hat{\mathcal{P}} \left(\prod_{i=1}^N \dd{x_i} \Psi_\kappa^* \Phi_\kappa \mathds{1}_\mathcal{R} \right).
\eeqa
Note that for the product of any two $N$-variables function $f $ and $g$
 one has 
 \beqa \label{eq:PermProd}
\hat{\mathcal{P}}(fg) = (\hat{\mathcal{P}}f)(\hat{\mathcal{P}}g).
\eeqa
Moreover, any wavefunction that follows exchange statistics parametrized by $\kappa$ must be an eigenstate of $\hat{\mathcal{P}}$
\beqa \label{eq:PermEigSt}
\hat{\mathcal{P}} \Psi_\kappa = p_\kappa \Psi_\kappa,
\eeqa
where $p_\kappa$ is an eigenvalue of magnitude 1, that only depends on $\kappa$ and the permutation $\mathcal{P}$. Properties (\ref{eq:PermProd}) and (\ref{eq:PermEigSt}) yield for the integral (\ref{eq:PermInt}):
\begin{align}
I(\mathcal{R}) & =\int_{\mathbb{R}^N} \prod_{i=1}^N \dd{x_i} \hat{\mathcal{P}}(\Psi_\kappa^*) \hat{\mathcal{P}}(\Phi_\kappa) \hat{\mathcal{P}}(\mathds{1}_\mathcal{R}) = \int_{\mathbb{R}^N} \prod_{i=1}^N \dd{x_i} \Psi_\kappa^* \underbrace{p_\kappa^* p_\kappa}_{=1} \Phi_\kappa \mathds{1}_{\mathcal{R}'}  \nonumber \\
& = \int_{\mathbb{R}^N} \prod_{i=1}^N \dd{x_i} \Psi_\kappa^* \Phi_\kappa \mathds{1}_{\mathcal{R}'} = I(\mathcal{R}').
\end{align}
The action of $\hat{\mathcal{P}}$ on $\mathds{1}_\mathcal{R}$ changes variables within the hierarchy dictated by $\mathcal{R}$; the integral is brought to a new sector $\mathcal{R}'$. All integrals are equal, independently from the selected sector, $\forall \mathcal{R},\mathcal{R}' \in S_N : I(\mathcal{R})=I(\mathcal{R}')$.

\section{On the derivation of the recursion relation} \label{App:4}

In this section, we give two different methods to derive the recurrence relation for $\Omega_N(\delta)$. Both methods take advantage of the recurrence structure of the permutation group, i.e., a permutation in the symmetric group $S_{N}$ can be viewed as the permutation of $S_{N-1}$ combined with an extra insertion of another element. The first method is based on a pure combinatoric argument while the second method makes use of an integral transform. In what follows, we shall not distinguish the difference between a sector and a permutation: a permutation $\mathcal{R}$ is identified with the sector where $x_{\mathcal{R}(1)}>x_{\mathcal{R}(2)}>\,\cdots>x_{\mathcal{R}(N)}$ whenever such identification is necessary.

\subsection{Combinatorial Method}
The statistical contribution 
\beqa
\Omega_N(\delta) := \frac{1}{N!} \sum_{\mathcal{R} \in S_N} \omega_\delta(\mathcal{R})
\eeqa
corresponds to the average of the phase factors $\omega_\delta(\mathcal{R})$ over all possible $N!$ sectors $\mathcal{R} \in S_N$. The phase factors are determined by the value $\hat{G}$ takes within the sector $\mathcal{R}$. Hence, for the sum over all phase factors:
\beqa
N! \, \Omega_N(\delta) = \sum_{\mathcal{R} \in S_N} \omega_\delta(\mathcal{R}) = \sum_{\mathcal{R} \in S_N} e^{-{\rm  i}\delta \hat{G}} \big|_\mathcal{R}.
\eeqa
We may single out the $N$-th variable from the generator $\hat{G}$
\beqa
N! \, \Omega_N(\delta) = \sum_{\mathcal{R} \in S_N} \left( \underbrace{e^{-{\rm  i}\delta \frac{1}{2} \sum_{i=1}^{N-1} {\rm sgn}(x_{iN})} \big|_\mathcal{R}}_{=A} \underbrace{e^{-{\rm  i}\delta \frac{1}{2} \sum_{i=1}^{N-1} {\rm sgn}(x_{ij})} \big|_\mathcal{R}}_{=B} \right).
\eeqa
The phase factor $A$ depends exclusively on the relative position of the $N$-th particle with respect to all other particles $i$, i.e., the position $n \in \{ 1,2,\dots,N \}$ within the hierarchy $x_{\mathcal{R}(1)}>x_{\mathcal{R}(2)}> \dots > x_{\mathcal{R}(N)}$ such that $x_{\mathcal{R}(n)}=x_N$. The factor $B$ is independent of the variable $x_N$ and behaves as the generator of a system with $N-1$ particles. By elimination of $x_N$ from $\mathcal{R}$ one obtains a remaining sector $\mathcal{R}'$
\beqa
\mathcal{R} & : & x_{\mathcal{R}(1)} > \dots > x_{\mathcal{R}(n-1)} > \overbrace{x_{\mathcal{R}(n)}}^{=x_N} > x_{\mathcal{R}(n+1)} > \dots > x_{\mathcal{R}(N)} \notag \\
& \downarrow & \label{eq:Elimination} \\
\mathcal{R}' & : & x_{\mathcal{R}(1)} > \dots > x_{\mathcal{R}(n-1)} > x_{\mathcal{R}(n+1)} > \dots > x_{\mathcal{R}(N)} \equiv x_{\mathcal{R}'(1)}>x_{\mathcal{R}'(2)}> \dots > x_{\mathcal{R}'(N-1)} \notag
\eeqa 
constructed from the variables $x_1,x_2,\dots,x_{N-1}$. Consider the tuple $(n,\mathcal{R'})$ that designates the sector $\mathcal{R}$ with $x_N$ in the $n$-th position and the remaining sector $\mathcal{R'}$. Since $S_N$ contains all possible $N$-particle sectors
\beqa
S_N = \bigcup_{n \in \{1,\dots,N\}, \, \mathcal{R}' \in S_{N-1}} \left(n,\mathcal{R}'\right),
\eeqa
where $S_{N-1}$ is the set of all $N-1$-particle sectors. This yields
\beqa
N! \, \Omega_N(\delta) & = & \sum_{n \in \{1,\dots,N\}, \, \mathcal{R}' \in S_{N-1}} \left( e^{-{\rm  i}\delta \frac{1}{2} \sum_{i=1}^{N-1} {\rm sgn}(x_{iN})} \big|_n \, e^{-{\rm  i}\delta \frac{1}{2} \sum_{i=1}^{N-1} {\rm sgn}(x_{ij})} \big|_{\mathcal{R}'} \right) \\
                       & = & \left(\sum_{n=1}^{N} e^{-{\rm  i}\delta \frac{1}{2} \sum_{i=1}^{N-1} {\rm sgn}(x_{iN})} \big|_{n} \right) \underbrace{\left( \sum_{\mathcal{R}' \in S_{N-1}} e^{-{\rm  i}\delta \frac{1}{2} \sum_{i=1}^{N-1} {\rm sgn}(x_{ij})} \big|_{\mathcal{R}'} \right)}_{= (N-1)! \, \Omega_{N-1}(\delta)}.
\eeqa
Trivially, 
\beqa
\left. \left( \sum_{i=1}^{N-1} {\rm sgn}(x_{iN}) \right) \right|_{n} = -(N-1-2(n-1)),
\eeqa
and we conclude the recursion relation
\beqa
\Omega_N(\delta) = \frac{1}{N} \sum_{n=1}^{N} e^{{\rm  i}\delta \frac{1}{2} (N-1-2(n-1)) } \Omega_{N-1}(\delta) = \frac{1}{N} \sum_{n=0}^{N-1} e^{{\rm  i} \delta \frac{1}{2} (N-1-2n) } \Omega_{N-1}(\delta),
\eeqa
which simplifies even further as the sum is geometric:
\beqa
\Omega_N(\delta) & = & \frac{1}{N} e^{{\rm  i} \delta \frac{1}{2} (N-1)} \underbrace{\sum_{n=0}^{N-1} e^{-{\rm  i} \delta n}}_{= \frac{1-e^{-{\rm  i} \delta N}}{1-e^{-{\rm  i} \delta}}} \Omega_{N-1}(\delta) \\
                 & = & \frac{1}{N} \frac{e^{{\rm  i}  \frac{1}{2} \delta N}-e^{-i  \frac{1}{2} \delta N}}{e^{{\rm  i} \frac{1}{2} \delta}-e^{-{\rm  i} \frac{1}{2} \delta}} \Omega_{N-1}(\delta) \\
								 & = & \frac{1}{N} \frac{\sin\left( \frac{N \delta}{2} \right)}{\sin\left( \frac{\delta}{2} \right)} \Omega_{N-1}(\delta).
\eeqa

\subsection{The method of integral transform}
We next provide an alternative derivation of $\Omega_N(\delta)$.
Our goal now is to transform the discrete sum over $\mathcal{R}$
in $\Omega_{N}(\delta)$ into an integral. To this end, we recover
the dependence on $\bm{x}=(x_{1},\,x_{2},\,\cdots x_{N})$ in $\Omega_{N}(\delta)$
and consider some integral transform of $\Omega_{N}(\delta,\,\bm{x})$
over the variables $\bm{x}$ with the kernel given by $\prod_{i}K_{i}(x_{i},\,s_{i})$.
So we find

\begin{align}
\tilde{\Omega}_{N}(\delta;\bm{s}) & \equiv\int_{\mathbb{U}^{N}}d^{N}\bm{x}\Omega_{N}(\delta;\bm{x})\prod_{i}K_{i}(x_{i},\,s_{i})\nonumber \\
 & =\frac{1}{N!}\int_{\mathbb{U}^{N}}d^{N}\bm{x}\sum_{\mathcal{R}\in S_{N}}e^{-\text{i}\frac{\kappa}{2}\sum_{1\le i<j\leq N}\text{sgn}\left(x_{i}-x_{j}\right)}\big|_{x_{\mathcal{R}(1)}>x_{\mathcal{R}(2)}>\cdots x_{\mathcal{R}(N)}}\prod_{i=1}^{N}K_{i}(x_{i},\,s_{i})\nonumber \\
 & =\frac{1}{N!}\int_{\mathbb{U}^{N}}d^{N}\bm{x}\sum_{\mathcal{R}\in S_{N}}e^{-\text{i}\frac{\kappa}{2}\sum_{1\le i<j\leq N}\text{sgn}\left(x_{i}-x_{j}\right)}\mathds{1}_{x_{\mathcal{R}(1)}>x_{\mathcal{R}(2)}>\,\cdots>x_{\mathcal{R}(N)}}\prod_{i=1}^{N}K_{i}(x_{i},\,s_{i})\nonumber \\
 & =\frac{1}{N!}\int_{\mathbb{U}^{N}}d^{N}\bm{x}e^{-\text{i}\frac{\kappa}{2}\sum_{1\le i<j\leq N}\text{sgn}\left(x_{i}-x_{j}\right)}\prod_{i=1}^{N}K_{i}(x_{i},\,s_{i}),\label{eq:Omg-hat}
\end{align}
where $\mathbb{U}^{N}\subset\mathbb{R}^{N}$ is the domain of the
integral transform. For example, for the Laplace transform, i.e., $K(x,\,s)=e^{-sx}$,
$\mathbb{U}=[0,\,+\infty]$.
We note that
\begin{equation}
\sum_{\mathcal{R}\in S_{N}}\mathds{1}_{x_{\mathcal{R}(1)}>x_{\mathcal{R}(2)}>\,\cdots>x_{\mathcal{R}(N)}}=\mathds{1}_{\mathbb{R}^{N}}
\end{equation}
and thus, it is also the identity in the domain of the integral transform
$\mathbb{U}$. 

Now, we are in a position to derive the recursive relation
between $\tilde{\Omega}_{N+1}(\delta;\,\bm{s})$ and $\tilde{\Omega}_{N}(\delta;\,\bm{s})$: They satisfy the following recursive relation
\begin{equation}
\tilde{\Omega}_{N+1}(\delta,\,\bm{s}_{N+1})=\frac{1}{N+1}\frac{\sin\left[(1+N)\kappa/2\right]}{\sin\left[\kappa/2\right]}\tilde{\Omega}_{N}(\delta;\,\bm{s})\int_{\mathbb{U}}dx_{N+1}K_{N+1}(x_{N+1},\,s_{N+1}).\label{eq:recursive-relation}
\end{equation}

\begin{proof}
It can be found that 

\begin{align}
\tilde{\Omega}_{N+1}(\delta,\,\bm{s}_{N+1}) & =\frac{1}{(N+1)!}\int_{\mathbb{U}^{N+1}}d^{N+1}\bm{x}e^{-\text{i}\frac{\kappa}{2}\sum_{1\le i<j\leq N+1}\text{sgn}\left(x_{i}-x_{j}\right)}\prod_{i=1}^{N+1}K_{i}(x_{i},\,s_{i})\nonumber \\
 & =\frac{1}{(N+1)!}\int_{\mathbb{U}^{N}}d^{N}\bm{x}e^{-\text{i}\frac{\kappa}{2}\sum_{1\le i<j\leq N}\text{sgn}\left(x_{i}-x_{j}\right)}\prod_{i=1}^{N}K_{i}(x_{i},\,s_{i}) \nonumber\\
 & \times\int_{\mathbb{U}}dx_{N+1}e^{-\text{i}\frac{\kappa}{2}\sum_{i=1}^{N}\text{sgn}\left(x_{i}-x_{N+1}\right)}K_{N+1}(x_{N+1},\,s_{N+1}).\label{eq:Omg-N+1+hat}
\end{align}
In addition, we note that 
\begin{equation}
\mathbb{U}^{N}=\sum_{\mathcal{R}}\mathbb{U}^{N}\cap\mathcal{R},
\end{equation}
where $\mathcal{R}$ denotes the sector $x_{\mathcal{R}(1)}>x_{\mathcal{R}(2)}>\,\cdots>x_{\mathcal{R}(N)}$.
Therefore we can rewrite Eq.~(\ref{eq:Omg-N+1+hat}) as follows
\begin{align}
\tilde{\Omega}_{N+1}(\delta,\,\bm{s}) & =\frac{1}{(N+1)!}\sum_{\mathcal{R}\in S_{N}}\int_{\mathbb{U}^{N}\cap I_{\mathcal{R}}}d^{N}\bm{x}e^{-\text{i}\frac{\kappa}{2}\sum_{1\le i<j\leq N}\text{sgn}\left(x_{i}-x_{j}\right)}\prod_{i=1}^{N}K_{i}(x_{i},\,s_{i}) \nonumber \\
& \times\int_{\mathbb{U}}dx_{N+1}e^{-\text{i}\frac{\kappa}{2}\sum_{i=1}^{N}\text{sgn}\left(x_{i}-x_{N+1}\right)}K_{N+1}(x_{N+1},\,s_{N+1}).
\end{align}
We denote 
\begin{equation}
\mathcal{R}_{k}:\,x_{\mathcal{R}(1)}>\cdots>x_{\mathcal{R}(k)}>x_{N+1}>x_{\mathcal{R}(k+1)}\cdots>x_{\mathcal{R}(N)},\,k=0,\,\cdots N.
\end{equation}
For every permutation $\mathcal{R}\in S_{N}$, one can always
separate the integral 
\begin{equation}
\int_{\mathbb{U}}dx_{N+1}=\sum_{k=0}^{N}\int_{\mathbb{U}\cap\mathcal{R}_{k}}dx_{N+1},
\end{equation}
which gives rise to
\begin{align}
\tilde{\Omega}_{N+1}(\delta,\,\bm{s}) & =\frac{1}{(N+1)!}\sum_{\mathcal{R}\in S_{N}}\int_{\mathbb{U}^{N}\cap\mathcal{R}}d^{N}\bm{x}e^{-\text{i}\frac{\kappa}{2}\sum_{1\le i<j\leq N}\text{sgn}\left(x_{i}-x_{j}\right)} \nonumber\\
& \times \prod_{i=1}^{N}K_{i}(x_{i},\,s_{i})\left[\sum_{k=0}^{N}\int_{\mathbb{U}\cap\mathcal{R}_{k}}dx_{N+1}e^{-\text{i}\frac{\kappa}{2}(N-2k)}K_{N+1}(x_{N+1},\,s_{N+1})\right]\label{eq:omg-Npls1}.
\end{align}
The following observation is key: 
\begin{equation}
\sum_{\mathcal{R}\in S_{N}}\int_{\mathbb{U}^{N}\cap\mathcal{R}}d^{N}\bm{x}\int_{\mathbb{U}\cap\mathcal{R}_{k}}dx_{N+1}=\int_{\mathbb{U}^{N}\cap(\cup_{\mathcal{R}}\mathcal{R})}d^{N}\bm{x}\int_{\mathbb{U}\cap(\cup_{\mathcal{R}}\mathcal{R}_{k})}dx_{N+1}=\int_{\mathbb{U}^{N}}d^{N}\bm{x}\int_{\mathbb{U}}dx_{N+1}.\label{eq:int-id}
\end{equation}
The meaning of Eq.~(\ref{eq:int-id}) is that if we sum over all
the permutations over $n$ coordinates $\bm{x}_{N}\equiv(x_{1},\,x_{2},\,\cdots x_{N})$,
then the constraint on the integration region vanishes, i.e., the
integration domain over $\bm{x}_{N}$ is left with $\mathbb{U}^{N}$
and the integration domain over $x_{N+1}$ is only $\mathbb{U}$.

With Eq.~(\ref{eq:int-id}), Eq.~(\ref{eq:omg-Npls1}) becomes
\begin{equation}
\tilde{\Omega}_{N+1}(\delta,\,\bm{s})=\sum_{k=0}^{N}\int_{\mathbb{U}^{N}}d^{N}\bm{x}e^{-\text{i}\frac{\kappa}{2}\sum_{1\le i<j\leq N}\text{sgn}\left(x_{i}-x_{j}\right)}\prod_{i=1}^{N}K_{i}(x_{i},\,s_{i})\left[\int_{\mathbb{U}}dx_{N+1}e^{-\text{i}\frac{\kappa}{2}(N-2k)}K_{N+1}(x_{N+1},\,s_{N+1})\right].
\end{equation}
Making use of the finite sum
\begin{equation}
e^{-\text{i}\frac{\kappa}{2}N}+e^{-\text{i}\frac{\kappa}{2}(N-2)}+\cdots e^{\text{i}\frac{\kappa}{2}(N-2)}+e^{\text{i}\frac{\kappa}{2}N}=\frac{\sin\left[(1+N)\kappa/2\right]}{\sin\left[\kappa/2\right]},
\end{equation}
we rewrite $\tilde{\Omega}_{N+1}(\delta,\,\bm{s})$ as 
\begin{equation}
\tilde{\Omega}_{N+1}(\delta,\,\bm{s})=\frac{\sin\left[(1+N)\kappa/2\right]}{\sin\left[\kappa/2\right]}\int_{\mathbb{U}^{N}}d^{N}\bm{x}e^{-\text{i}\frac{\kappa}{2}\sum_{1\le i<j\leq N}\text{sgn}\left(x_{i}-x_{j}\right)}\prod_{i=1}^{N}K_{i}(x_{i},\,s_{i})\int_{\mathbb{U}}dx_{N+1}K_{N+1}(x_{N+1},\,s_{N+1})],
\end{equation}
which concludes the proof.
\end{proof}
Applying the inverse transform on both sides of Eq.~(\ref{eq:recursive-relation}),
we find
\begin{equation}
\Omega_{N+1}(\delta,\,\bm{x})=\frac{\sin\left[(1+N)\kappa/2\right]}{\sin\left[\kappa/2\right]}\Omega_{N}(\delta,\,\bm{x})\label{eq:recursive-Omg}.
\end{equation}

\section{The summation expression of $\Omega_N$ and combinatorics}

Appendix~\ref{App:4} determines an exact closed-form expression for $\Omega_N(\delta)$, as it transforms the summation into a product, from which the zeros can be determined. The summation expression is rather cumbersome as it raises a combinatorics problem: the sum goes over $N!$ sectors, but the spectrum $\hat{G}$ only allows for $N(N+1)/2+1$ distinct eigenvalues, so the sum must be highly degenerate. Hence,
\beqa
\Omega_N(\delta) = \frac{1}{N!} \sum_{n=0}^{N(N-1)/2} a(n,N) \omega_\delta(n), \textrm{ where } \omega_\delta(n) = e^{-i \delta g_n} \textrm{ and } g_n = - \frac{N(N-1)}{4} +  n,
\eeqa
which is obtained from Eq. (\ref{eq:SpecG}). The combinatorics problem involves determining the number $a(n,N)$ of distinct sectors $\mathcal{R} \in S_N$ that lead to a same factor $\omega_\delta(n)$ or eigenvalue $g_n$, equivalently. Note that the problem is not solved by choosing a certain eigenvalue $g_n$, determining the number $n$ of signs sgn$(x_{ij})$ that must be positive and selecting $a(n,N)$ as the number of ways to choose $n$ positive signs within $N(N-1)/2$ (this leads to binomial coefficients), as not all combination of sgn$(x_{ij})$ are legal: For instance sgn$(x_{ij}) = +1$, sgn$(x_{jk}) = +1$ and sgn$(x_{ik}) = -1$ is contradictory. 

Nonetheless, there is a solution that follows from recursion alike Appendix.~\ref{App:4}: Let be $\mathcal{R}' \in S_{N-1}$ a sector of variables $x_1,x_2,\dots,x_{N-1}$, for which we know there are $n'$ signs positive. From $\mathcal{R}'$ a sector $\mathcal{R} \in S_N$ with $n=n'+k,k\in\{0,1,\dots,N-1\}$ positive signs can be constructed through the bijective map: 
\beqa
\mathcal{M}_k : \left\{ \mathcal{R}' \in S_{N-1} \big| \hat{G}^{(N-1)}\mathds{1}_{\mathcal{R}'} = g_{n'}^{(N-1)}\mathds{1}_{\mathcal{R}'} \right\}  \rightarrow \left\{ \mathcal{R} \in S_N \big| \hat{G}^{(N)}\mathds{1}_{\mathcal{R}} = g_{n'+k}^{(N)}\mathds{1}_{\mathcal{R}}, x_N = x_{\mathcal{R}(k+1)} \right\} \label{eq:Map}
\eeqa
\begin{align}
& x_{\mathcal{R}'(1)} > \dots > x_{\mathcal{R}'(k)} > x_{\mathcal{R}'(k+1)} > \dots > x_{\mathcal{R}(N-1)}  \nonumber \\
&\mapsto \underbrace{ x_{\mathcal{R}'(1)} > \dots > x_{\mathcal{R}'(k)} > x_N > x_{\mathcal{R}'(k+1)} > \dots > x_{\mathcal{R}'(N-1)}}_{\equiv x_{\mathcal{R}(1)}>x_{\mathcal{R}(2)}> \dots > x_{\mathcal{R}(N)}},
\end{align}
as the $k$ signs sgn$(x_{\mathcal{R}(1)N})$ up to sgn$(x_{\mathcal{R}(k)N})$ are positive in addition to the $n'$ that are inherited unchanged from $\mathcal{R}'$. $\hat{G}^{(N)}$ and $g^{(N)}$ indicate that the generator respectively the corresponding eigenvalue are $N$-particle. $\mathcal{M}_k$ corresponds to slipping the variable $x_N$ into the position $k+1$, and can be understood analogously to the inverse of operation (\ref{eq:Elimination}). Hence, we determine a one-to-one correspondence between sectors $\mathcal{R} \in S_N$ that fulfil the condition $\hat{G}\mathds{1}_{\mathcal{R}} = g_{n}\mathds{1}_{\mathcal{R}}$ and sectors $\mathcal{R}' \in S_{N-1}$. Furthermore,
\beqa
\underbrace{\left\{ \mathcal{R} \in S_N \big| \hat{G}^{(N)} \mathds{1}_\mathcal{R}=g_n^{(N)}\mathds{1}_{\mathcal{R}'} \right\}}_{=S_N^{g_n}} = \bigcup_{k=0}^{N-1} \underbrace{\left\{ \mathcal{R} \in S_N \big| \hat{G}^{(N)}\mathds{1}_{\mathcal{R}} = g_n^{(N)}\mathds{1}_{\mathcal{R}}, x_N = x_{\mathcal{R}(k+1)} \right\}}_{=S_N^{g_n,k}} \label{eq:Union},
\eeqa
and $\forall N \in \mathbb{Z}_+, \forall g_n,g_m \in {\it Sp}(\hat{G})$ s.t. $g_n \neq g_m, \forall k,l \in \{0,1,\dots,N-1\} $ s.t. $k \neq l$:
\beqa
S_N^{g_n}\cap S_N^{g_m} = \varnothing = S_N^{g_n,k}\cap S_N^{g_n,l}.
\eeqa
Thus, the recurrence relation follows from Eq. (\ref{eq:Map}) and Eq. (\ref{eq:Union})
\beqa
a(n,N) = \textrm{Card}\left( S_N^{g_n} \right) = \sum_{k=0}^{N-1} \textrm{Card}\left(S_N^{g_n,k}\right) = \sum_{k=0}^{N-1} a(n-k,N-1) = \sum_{k=n+1-N}^{n} a(k,N-1),
\eeqa
where $\forall N \in \mathbb{Z}_+$
\beqa
a(n<0,N) = 0 = a\left(n>\frac{N(N-1)}{2},N\right),
\eeqa
since the corresponding eigenvalues $g_n$ do not exist. Clearly, $a(0,1)=1$ initiates the recurrence and hence one can construct the number table in Fig. \ref{Fig5MackelPile}, which presents a few interesting properties:
\begin{figure}[h!]
\centering
\begin{tikzpicture}
\node (image) at (-2,0) {\includegraphics[angle = 90, scale = .7]{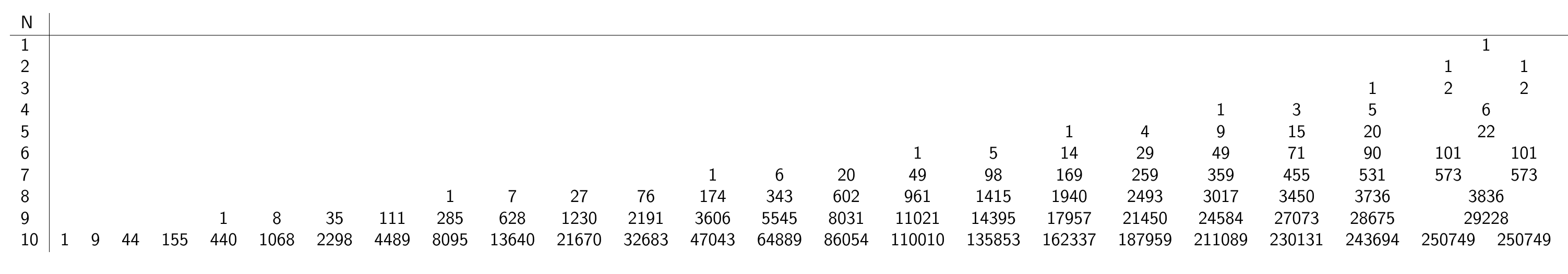}};
\node (image) at (2,0) {\includegraphics[angle = 90, scale = .7]{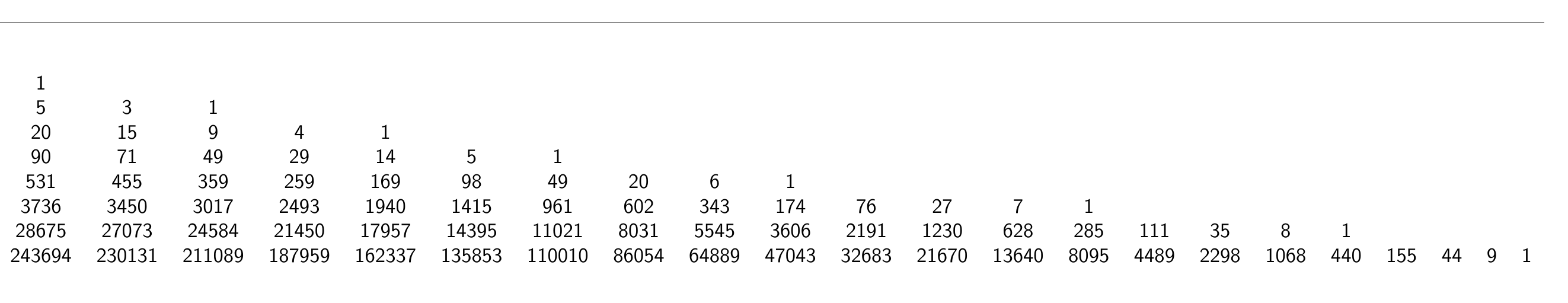}};
\draw[->] (2,9.8) to [out=150,in=30] (-2,10.8);
\end{tikzpicture}
\caption{{\bf Pascal-esk representation of the eigenvalue-degeneracy of the generator.} Number $a(n,N)$ of eigenvalues $g_n$, where $n$ evolves from $0$ to $N(N-1)/2$ in steps of $1$ going left in a row.}
\label{Fig5MackelPile}
\end{figure}
\begin{itemize}
	\item The table is symmetric over its center
	\beqa
	a(n,N) = a\left( \frac{N(N-1)}{2}-n, N \right),
	\eeqa
	as for every sector $\mathcal{R}$ there exists a conjugate sector $\bar{\mathcal{R}}$ defined by $x_{\mathcal{R}(N)} > x_{\mathcal{R}(N-1)} > \dots > x_{\mathcal{R}(1)}$ that has eigenvalue $g_{N(N-1)/2-n}$.
	\item The recurrence can be understood graphically, with each element of the $N$-th row being equal to the sum of the $N$ elements above from the $N-1$-th row. Regarding the recurrence, all sectors $\mathcal{R} \in S_N$ with eigenvalue $g_n^{(N)}$ can be constructed uniquely from the sectors $\mathcal{R}' \in S_{N-1}$ with eigenvalues $g_{n-k},k\in\{0,1,\dots,N-1\}$, by slipping in the $N$-th position variable.
	\item The table predicts for any $N$ a sum of numbers that is equal to $N!$ - which is quite curious in itself but may be interesting in number theory. Intuitively, this must be the case since there are in total $N!$, whereas it can be shown explicitly from the recurrence:
	\begin{itemize}
		\item \underline{Initial step}: $a(0,1) = 1 = 1!$
		\item \underline{Recurrence step}: 
		\beqa
		\sum_{n=0}^{N(N-1)/2} a(n,N) = \sum_{n=0}^{N(N-1)/2} \sum_{k=n+1-N}^{n} a(k,N-1) = N \underbrace{\sum_{i=0}^{(N-1)(N-2)/2} a(i,N)}_{=(N-1)! \textrm{ by hypothesis}} = N!
		\eeqa
	\end{itemize}
\end{itemize}
Aside from being also the solution to a combinatorics problem, the table shown in Fig. \ref{Fig5MackelPile}  is analogous to Pascal's triangle in terms of the above properties. Pascal's triangle is symmetric as well, the sum of the elements of the $N$-th row equates to $2^N$, and any element is the sum of the two above. 

The recurrence relation can be used to determine the explicit expressions of $a(n,N)$. For the first few values of $n$, these are:
\beqa
a(0,N) & = & 1,\\ 
a(1,N) & = & N-1,  \\
a(2,N) & = & \frac{(N+1)(N-2)}{2},  \\
a(3,N) & = & \frac{N(N^2-7)}{6},  \\
a(4,N) & = & \frac{N(N+1)(N^2+N-14)}{24},  \\ 
a(5,N) & = & \frac{(N-1)(N+6)(N^3-9N-20)}{120}.
\eeqa
 
 \section{Quantum speed limits under shifts of the statistical parameter \label{QSL}}
 
 \subsection{Mandelstam-Tamm uncertainty relation and speed limit}
 Given a pure state $|\Psi_\kappa\ra$, the Heisenberg uncertainty relation reads
 \beqa
 \Delta O \Delta G\geq\frac{1}{2}\left| \la \Psi_\kappa|[\hat{O}_H(\kappa),\hat{G}]|\Psi_\kappa\ra\right|,
 \eeqa
 which, using the Heisenberg equation of motion, yields

  \beqa
 \Delta O \Delta G\geq\frac{1}{2}\left| \frac{d}{d\kappa}\la \Psi_\kappa|\hat{O}_H(0)|\Psi_\kappa\ra\right|.
 \eeqa
 
 The characteristic shift for the mean $\la \Psi_\kappa|\hat{O}_H(0)|\Psi_\kappa\ra$ to vary by a value $ \Delta O $ is estimated as
 \beqa
 \kappa_O:=\frac{ \Delta O }{\left| \frac{d}{d\kappa}\la \Psi_\kappa|\hat{O}_H(0)|\Psi_\kappa\ra\right|}.
 \eeqa
 With it, one arrives at the Mandelstam-Tamm uncertainty relation
 \beqa
  \kappa_O\Delta G\geq\frac{1}{2},
 \eeqa
 which provides a lower bound to $  \kappa_O$ in terms of the variance of the generator $\Delta G$.
 The MT QSL can be found as an orthogonalization bound by choosing  $O$ to be the projector on to the initial state, as in the original study by Mandelstam and Tamm \cite{Mandelstam45}.

 \subsection{Margolus-Levitin bound}
 We next aim at finding a lower bound for the shift of $\kappa$ required for the orthogonalization of a state $\Psi$.
 For identical particles, any many-particle state $\Psi\in\mathcal{L}^2(\mathbb{R}^N)$ as support in the spectrum of $\hat{G}$.
 The action of the shift of the statistical parameter $\kappa$ generated by $\hat{G}$ on the state $\Psi_0$ is described by
 \beqa
 \Psi_\kappa=\sum_{\mathcal{R} \in S_N} \omega_\delta(\mathcal{R})\,\mathds{1}_\mathcal{R}|\Psi_0\ra=\sum_{\mathcal{R} \in S_N} e^{-{\rm  i}g_\mathcal{R}}\,\mathds{1}_\mathcal{R}|\Psi_0\ra.
 \eeqa
 Along the $\kappa$-flow, the survival amplitude of the initial state is given by the overlap 
 \beqa
\la\Psi_0|\Psi_\kappa \ra=\sum_{\mathcal{R} \in S_N} e^{-{\rm  i}g_\mathcal{R}}p_\mathcal{R},
 \eeqa
 where $p_\mathcal{R}=\la\Psi_0|\mathbb{I}_\mathcal{R}|\Psi_0\ra$.
 By permutation symmetry, it can  be shown that 
 \beqa
 p_\mathcal{R}=\frac{1}{N!}.
 \eeqa
 We can next proceed by tweaking slightly the derivation by Margolus and Levitin \cite{Margolus98}, noting that the real part of the survival amplitude
 \beqa
 {\rm Re}[e^{-{\rm  i}g_-}\la\Psi_0|\Psi_\kappa\ra]&=&\frac{1}{N!}\sum_{\mathcal{R} \in S_N} \cos[(g_\mathcal{R}-g_-)\kappa]\\
 &\geq &\frac{1}{N!}\sum_{\mathcal{R} \in S_N} \left(1-\frac{2}{\pi}(g_\mathcal{R}-g_-)\kappa-\frac{2}{\pi}\sin[(g_\mathcal{R}-g_-)\kappa]\right)\\
 &=&1+\frac{2}{\pi}g_-\kappa+\frac{2}{\pi}{\rm Im}[e^{-{\rm  i}g_-}\la\Psi_0|\Psi_\kappa\ra],
 \eeqa
 where we have made use of the fact that $\la \hat{G}\ra=\frac{1}{N!}\sum_{\mathcal{R} \in S_N}g_\mathcal{R}=0$.
 Imposing $\la\Psi_0|\Psi_\kappa\ra=0$, it follows that the required shift is lower bounded by
 \beqa
 \kappa\geq \kappa_{\rm ML}=-\frac{\pi}{2g_-}=+\frac{2\pi}{N(N-1)}.
 \eeqa

\section{Universal Orthogonality catastrophe under shifts of the statistical parameter}

Note that the statistical contribution $\Omega_N(\delta)$ is the moment generating function to the generator $\hat{G}$
\beqa
\Omega_N(\delta) = \langle \Psi_\kappa | e^{-{\rm  i}\delta \hat{G}} | \Psi_\kappa \rangle.
\eeqa
Hence, we may consider looking at the cumulant generating function
\beqa
\sum_{n=1}^{+\infty} \frac{(-{\rm  i} \delta)^n}{n!} \langle  \hat{G}^n \rangle_c = \ln(\Omega_N(\delta)) = \sum_{n=2}^N \ln\left( \frac{\sin\left( \frac{n \delta}{2} \right)}{n \sin\left( \frac{\delta}{2} \right)} \right),
\eeqa
where the cumulants are given by 
\beqa
\langle \hat{G}^n \rangle_c = \frac{1}{(-{\rm  i})^n} \lim_{\delta \to 0} \frac{\dd{}^n}{\dd{\delta}^n} \ln\left( \Omega_N(\delta) \right).
\eeqa 
Since $\Omega_N(\delta)$ is an even function, all odd cumulants must be zero: $\langle \hat{G}^{\rm odd} \rangle_c =0$. The first few even cumulants are:
\beqa
\langle \hat{G}^2 \rangle_c & = & \sum_{n=2}^N \frac{1}{12} (n^2-1) \\
                            & = & \frac{N(2N^2+3N-5)}{72}, \\
\langle \hat{G}^4 \rangle_c & = & - \sum_{n=2}^N \frac{1}{120} (n^4-1) \\
                            & = & -\frac{N(6N^4+15N^3+10N^2-31)}{3600}, \\
\langle \hat{G}^6 \rangle_c & = & \sum_{n=2}^N \frac{1}{252} (n^6-1) \\
                            & = & \frac{N(6N^6+21N^5+21N^4-7N^2-41)}{10584}, \\
\langle \hat{G}^8 \rangle_c & = & - \sum_{n=2}^N \frac{1}{240} (n^8-1) \\
                            & = & - \frac{N(10N^8+45N^7+60N^6-42N^4+20N^2-93)}{21600}, \\
\langle \hat{G}^{10} \rangle_c & = & \sum_{n=2}^N \frac{1}{132} (n^{10}-1) \\
                               & = & \frac{N(6N^{10}+33N^9+55N^8-66N^6+66N^4-33N^2-61)}{8712}.
\eeqa
The pattern is persistent: The $k$-th derivative of the cumulant generating function can be written as
\beqa
\frac{\dd{}^k}{\dd{\delta}^k} \ln(\Omega_N(\delta)) = 2^{-k} \sum_{n=2}^N \lim_{\delta \to 0} \left[ n^k \left. \left(\frac{\dd{}^k}{\dd{t}^k} \ln(\sin(t)) \right) \right|_{t=\frac{n \delta}{2}} - \left. \left(\frac{\dd{}^k}{\dd{t}^k} \ln(\sin(t)) \right) \right|_{t=\frac{\delta}{2}} \right],
\eeqa
so that the lead-term of the $k$-th cumulant in $N$ must be to the power $k+1$. At large $N$ all even numbered cumulants can be written as
\beqa
\langle \hat{G}^n \rangle_c \approx \alpha_n N^{n+1},
\eeqa
where $\forall n$,  $\alpha_n \in \mathbb{R}$. 
The cumulant expansion allows to write the statistical contribution $\forall \delta \in \left[-\frac{2 \pi}{N},\frac{2 \pi}{N} \right]$ as 
\beqa
\Omega_N(\delta) = \exp[\sum_{{\rm even \;} n=2}^{+\infty} \frac{(-{\rm  i} \delta)^n}{n!} \langle  \hat{G}^n \rangle_c ],
\eeqa
which allows for approximation by truncating higher-order terms.
Consider exclusively small shifts in statistics, i.e., $\delta$ restricted to the interval
\beqa
I_{<t} = \left[- \sqrt[4]{\frac{\ln(1+t)}{|\alpha_4| N^5}}, \sqrt[4]{\frac{\ln(1+t)}{|\alpha_4| N^5}} \right],
\eeqa
where $t >0$ is an arbitrarily chosen threshold. Then, 
$\forall \delta \in I_{<t}, \forall n > 4$ the terms of the cumulant expansion dies out at large $N$:
\beqa
\alpha_n \delta^n N^{n+1} \big|_{I_{<t}} \leq \alpha_n \left( \sqrt[4]{\frac{\ln(1+t)}{|\alpha_4| N^5}} \right)^n N^{n+1} \propto N^{-\frac{n}{4}+1} \stackrel{N \to +\infty}{\rightarrow} 0.
\eeqa
Furthermore, for the term $n=4$
\beqa
\alpha_4 \kappa^4 N^5 \big|_{I_{<t}} \leq \frac{\alpha_4}{|\alpha_4|} \ln(1+t) = - \ln(1+t),
\eeqa 
so that the relative error made by approximating $\Omega_N(\delta)$ with 
\beqa
\mathcal{G}_N(\delta) = e^{-\frac{\delta^2}{2} \frac{N (2N^2+3N-5)}{72}} = e^{-\frac{\delta^2}{2 \sigma^2}}
\eeqa 
is upper-bounded:
\beqa
\max_{I_{<t}} \left| \frac{\Omega_N(\delta)-\mathcal{G}_N(\delta)}{\Omega_N(\delta)} \right| = \left| 1 - e^{\ln(1+t)} \right| = t.
\eeqa
Consequently, $I_{<t}$ is the interval within which the error of approximation is smaller than the threshold $t$; by choosing the appropriate interval $I_{<t}$, the error can be rendered arbitrarily small. In addition, this interval contracts slower than any interval $I_\sigma = [-b \sigma, b \sigma]$ with ($b>0$) that scales with the variance of the Gaussian approximation, so that in the thermodynamic limit $\forall t: I_\sigma \subseteq I_{<t}$, and the approximation for all relevant behavior must be exact. Even better approximations may be obtained by holding on to additional lower-order cumulants.

\medskip
\bibliographystyle{unsrtnat}
 \bibliography{QAlchemy_lib} 
 
\end{document}